\documentclass[useAMS,usenatbib]{mnras}

\usepackage{url,times,hyperref,graphicx,amsmath,amsfonts,amssymb,aas_macros,color,epsfig,epstopdf}
\usepackage{dblfloatfix} 

\newcommand{\hMpc}{{\ifmmode{h^{-1}{\rm Mpc}}\else{$h^{-1}$Mpc}\fi}}
\newcommand{\hkpc}{{\ifmmode{h^{-1}{\rm kpc}}\else{$h^{-1}$kpc}\fi}}
\newcommand{\hMsun}{{\ifmmode{h^{-1}{\rm {M_{\odot}}}}\else{$h^{-1}{\rm{M_{\odot}}}$}\fi}}
\newcommand{\ltsima}{$\; \buildrel < \over \sim \;$}
\newcommand{\gtsima}{$\; \buildrel > \over \sim \;$}
\newcommand{\lsim}{\lower.5ex\hbox{\ltsima}}
\newcommand{\gsim}{\lower.5ex\hbox{\gtsima}}

\def\lesssim{\mathrel{\hbox{\rlap{\hbox{\lower4pt\hbox{$\sim$}}}\hbox{$<$}}}}
\def\gtrsim{\mathrel{\hbox{\rlap{\hbox{\lower4pt\hbox{$\sim$}}}\hbox{$>$}}}}

\newcommand{\beq}{\begin{equation}}
\newcommand{\eeq}{\end{equation}}
\def\beqa{\begin{eqnarray}}
\def\eeqa{\end{eqnarray}}
\def\hMpc{$h^{-1}\,{\rm Mpc}$}
\def\hkpc{$h^{-1}\,{\rm kpc}$}

\def\head{
 \vbox to 0pt{\vss
                   \hbox to 0pt{\hskip 440pt\rm LA-UR-10-07069\hss}
                  \vskip 25pt}}

\title[Alpha abundances of the Milky Way's halo and old thick disc]
{Explaining the chemical trajectories of accreted and in-situ 
halo stars of the Milky Way}
\author[Brook et al.]
       {Chris B. Brook$^{1,2}$, Daisuke Kawata$^{3}$, Brad K. Gibson$^{4}$, Carme Gallart$^{2,1}$, Andr\'{e}s Vicente$^{2,1}$\\
$^{1}$Universidad de La Laguna. Avda. Astrof\'{i}sico Fco. S\'{a}nchez, La Laguna, Tenerife, Spain\\
$^{2}$Instituto de Astrof\'{i}sica de Canarias, Calle Via L\'{a}ctea s/n, E-38206 La Laguna, Tenerife, Spain\\
$^{3}$Mullard Space Science Laboratory, University College London, Holmbury St. Mary, Dorking, Surrey, RH5 6NT, UK\\
$^{4}$E. A. Milne Centre for Astrophysics, University of Hull, Hull, HU6 7RX, United Kingdom\\
}

\begin{document}

\date{Accepted xxxx. Received xxxx; in original form xxxx}

\pagerange{\pageref{firstpage}--\pageref{lastpage}} \pubyear{2020}

\maketitle

\label{firstpage}

\begin{abstract}
The Milky Way underwent its last significant merger ten billion years 
ago, when the Gaia-Enceladus-Sausage (GES) was accreted. 
Accreted GES stars and progenitor stars born prior to the merger make up the bulk of the inner halo. Even 
though these two main populations of halo stars have 
similar {\it durations} of star formation prior to their merger, 
they differ in [$\alpha$/Fe]-[Fe/H] space, with the GES population 
bending to lower [$\alpha$/Fe] at a relatively low value of [Fe/H].  We 
use cosmological simulations of a 'Milky Way' to argue that the 
different tracks of the halo stars through the [$\alpha$/Fe]-[Fe/H] 
plane are due to a difference in their star formation history and 
efficiency, with the lower mass GES having its low and constant star 
formation regulated by feedback whilst the higher mass main progenitor 
has a higher star formation rate prior to the merger. 
The lower star formation efficiency of GES leads to lower gas pollution 
levels, pushing [$\alpha$/Fe]-[Fe/H] tracks to the left. In addition, the increasing star formation rate maintains a 
higher relative contribution of Type~II SNe to Type~Ia SNe for the main 
progenitor population that formed during the same time period, 
thus maintaining a relatively high [$\alpha$/Fe]. Thus the 
different positions of the downturns in the [$\alpha$/Fe]-[Fe/H] plane 
for the GES stars are not reflective of different star formation 
durations, but instead reflect different star formation efficiencies.
\end{abstract}

\noindent
\begin{keywords}
galaxies: dwarf - evolution - formation - haloes - clusters
\end{keywords}

\section{Introduction} \label{sec:introduction}
The Galactic halo can be defined kinematically as stars with tangential 
velocities greater than 200~km~s$^{-1}$ relative to the local standard 
of rest. So defined, the local stellar halo (within $\sim$2~kpc at 
least) is largely comprised of two populations. One population has 
relatively low eccentricities whilst the other  has relatively 
large radial motions and high eccentricities \cite[e.g.][]{mackereth18,amarante20}. 
This latter population was first identified by its kinematics using 
Hipparcos data \citep{chibabeers00} and has long been considered to have 
originated from an accreted satellite \citep{brook03}.

Recent studies confirmed that this phase-space structure is a major 
contributor to the stellar halo \citep{helmi18,belokurov18,iorio19}, and 
originates in the most significant accretion event in the Milky Way's 
history. The accreted galaxy that contributed stars to this structure is 
now referred to variously as Gaia-Enceladus \citep{helmi18} and the Gaia 
sausage \citep{belokurov18}. We refer to it as the
Gaia-Enceladus-Sausage (GES) in what follows. A second, smaller accreted 
galaxy, dubbed SEQUOIA, has also been identified \citep{myeong19}.

The presence of two kinematically and chemically distinct halo 
populations has been long discussed. A difference in age of 2-3 Gyr had 
previously been inferred, with the low-alpha, accreted population being 
assigned younger ages \citep[e.g.][]{Schuster+Moreno+Nissen+12}. These 
two dominant populations of inner halo stars also separate in their 
Hertzsprung-Russell diagram (HRD) \citep{Gaia+Babusiaux+18,haywood18}, 
with the stars originating in GES forming a blue sequence, whilst the 
other population forms a red sequence.  However, using robust techniques 
of CMD-fitting to derive age distributions of stellar 
populations, which had previously been used on nearby dwarf galaxies, 
\cite{gallart19} found that the two inner halo populations are actually 
coeval.\footnote{To be clear, our analysis is focussed upon the inner
halo population; age gradients from the inner to the outer halo are 
the focus of complementary works such as \cite{carollo18}.}

Adopting these findings regarding ages points to a particular formation sequence of the 
Milky Way: the red sequence halo stars were heated from the more massive 
main progenitor of the Milky Way (MW$_{prog}$) during the accretion of 
GES and thus can be considered the 'in situ halo' using the terminology 
of \cite{zolotov09}, whilst the blue sequence halo stars belonged to the 
accreted GES. The coeval ages of the two halo populations implies that 
no stars that formed in the Milky Way's main progenitor after this 
accretion event were heated to halo-like kinematics, meaning that the 
GES accretion can be considered the last significant (possibly major) 
merger event in the Milky Way's history.
 
Using the {\it Gaia} second data release \citep[DR2][]{Gaia+Brown+Vallenari+18} 
and spectroscopic data from LAMOST \citep{Zhao+Zhao+Chu+12} and GALAH 
\citep{Bunder+Asplund+Duong+18}, \cite{gallart19} showed that the thick 
disc continued to form most of its stars after the GES event and, thus, 
is dominated by stars younger than these two halo 
populations (although with a tail to older stars). 
The thick disc stars fall on the 
same sequence of the CMD as the red sequence halo stars. Thus, the stars 
in the red sequence were formed 'in situ' in the Milky Way's main 
progenitor, before, during, and after the GES merger event. However, it 
was only during the GES event that some main progenitor stars were 
heated enough to be classified as (in situ) halo stars. In the following 
we refer to both  'in situ halo' and thick disc stars, i.e. old stars on 
the red sequence of the CMD, as 'main progenitor stars'.  'Main 
progenitor stars' in this context are not taken to include the thick 
disc stars that formed after the merger, nor the thin disc stars, even 
though they clearly also form within the same galactic structure. While 
the later forming thick disc stars still fall in the same red sequence 
as the so-defined 'main progenitor stars', they tend to be even redder 
than the red sequence of the kinematically-selected halo.
 
Using complementary arguments based on a combination of kinematic 
information from {\it Gaia}~DR2 and detailed chemical abundances from 
the SDSS/APOGEE \citep{Holtzman+Hasselquist+Shetrone+18} data, 
\cite{dimatteo19} also conclude that the stars in the red halo sequence 
were formed within the main progenitor. They refer to this population as 
'thick disc' stars, due to their chemical similarity. Any difference 
between these results and that of \cite{gallart19} regarding this 
population of stars is purely semantic. Both studies identify these 
stars as being formed within the main progenitor, with 
\cite{haywood18,dimatteo19} calling them the high velocity tail of the 
thick disc, whilst \cite{gallart19} follows the terminology of 
\cite{zolotov09} in calling them 'in situ halo' stars. Results from FIRE 
simulations \citep{bonaca17} also support the notion that the highest 
metallicity halo stars were formed 'in situ'. What is important to note 
is that the age distributions inferred by \citet{gallart19} imply that 
1) only main progenitor stars that formed {\it before} (and perhaps 
during) the merger with GES gain such high velocities, and 2) thick disc 
stars were forming prior to, during, and after this merger. Further 
support for these conclusions comes from abundances of high velocity 
stars uncovered by Skymapper, and ages derived using isochrone fitting 
techniques \citep{sahlholdt19}.
 
A key component of the \cite{dimatteo19} study is the use 
of $\alpha$-elements, primarily formed in Type~II supernovae 
(SNeII), and their comparison with abundances of Fe, largely formed 
in Type~Ia supernovae (SNeIa) and thus delayed in time compared to 
$\alpha$-element production.  In particular, the GES stars fall in a 
different region of the [Fe/H] vs [O/Fe] plane than the old main 
progenitor (thick disc and 'in situ' halo) stars, with the GES stars 
bending downward toward low values of [O/Fe] at a lower value of [Fe/H] 
than the main progenitor stars.
 
Earlier, \cite{nissen10} had used kinematics and [O/Fe] to argue that a 
sample of stars that today we know are part of the "blue sequence" halo 
stars were likely accreted, whilst stars falling in the red sequence 
were either heated from the early forming main progenitor or formed "as 
the first stars during the collapse of a proto-Galactic gas cloud". 
\cite{haywood18} also concluded that the 
'blue sequence' halo stars were likely accreted, and agreed with the 
scenario whereby the red sequence halo stars were heated from the main 
progenitor - in particular, that they were heated from the early-forming 
thick disc.  \cite{gallart19} contributed to this
scenario by showing that the red sequence 'in situ' halo stars were 
heated from the disc {\it by the same event} that incorporated the blue 
sequence stars into the inner halo - i.e., by the accretion of the GES. 
The age information provided by the latter has allowed a clear
picture of the Milky Way's formation to emerge.
 
Efforts have been made to use hydrodynamical simulations to constrain 
the mass of the GES galaxy using metallicities and/or $\alpha$-abundances 
to infer the mass of the accreted 
galaxies that contributed to the formation of the inner halo 
\citep{robertson05,font06,zolotov10,tissera13,fattahi19,mackereth18,fernandez19,vincenzo19}.
Our simulations support the notion 
that the chemical abundance patterns 
of the halo stars reflect the fact 
that a relatively massive accretion event occurred.  In this paper, we 
study the {\it origin} of the difference in chemical 
abundances between halo stars accreted from the GES and those stars born in 
the main progenitor (and heated by the interaction with the GES).
 
Previously, \cite{nissen10}, following from \cite{gilmore98}, suggested 
that the origin of the difference is the {\it duration} of star 
formation, with the low [$\alpha$/Fe] population forming over a longer time 
such that they are more enriched by SNeIa. This fit with the previously 
held idea that the GES stars were {\it younger} than the main progenitor 
halo stars.  \cite{haywood18} invoked the difference in star formation 
{\it efficiency} for explaining the two tracks in the [Fe/H] vs [O/Fe] plane,
without putting constraints on the differences in the duration of star 
formation. \cite{fernandez19} suggest that it is a combination of both 
differences in the duration and the intensity of star formation, 
combined with differences in the initial mass functions \citep[see 
also][]{kobayashi14} of the two merging galaxies, that explains the 
different chemical evolution tracks.
  
Here, we leverage the result of \cite{gallart19} to place new 
constraints on the origin of these tracks in abundance space. 
\cite{gallart19} show that the two halo populations were formed over 
essentially the same period. Thus, the different tracks in the [Fe/H] vs 
[$\alpha$/Fe] plane are {\it not due to a longer time-span of star 
formation} in the GES galaxy compared to the population of halo stars 
that formed in the main progenitor.

We explore a cosmological simulation of a Milky Way analogue which has a 
merger history that resembles that of the Galaxy, with an early 
last significant merger. In the simulated galaxy, the accreted galaxy 
and main progenitor have similar [Fe/H] vs [$\alpha$/Fe] distributions 
as observed in the blue and red sequences of the Milky Way local stellar 
halo. We identify the reason for the different tracks as being a 
combination of the different star formation efficiencies and the star 
formation history before the merger.

\begin{figure}
\includegraphics[width=3.4in,height=2.5in]{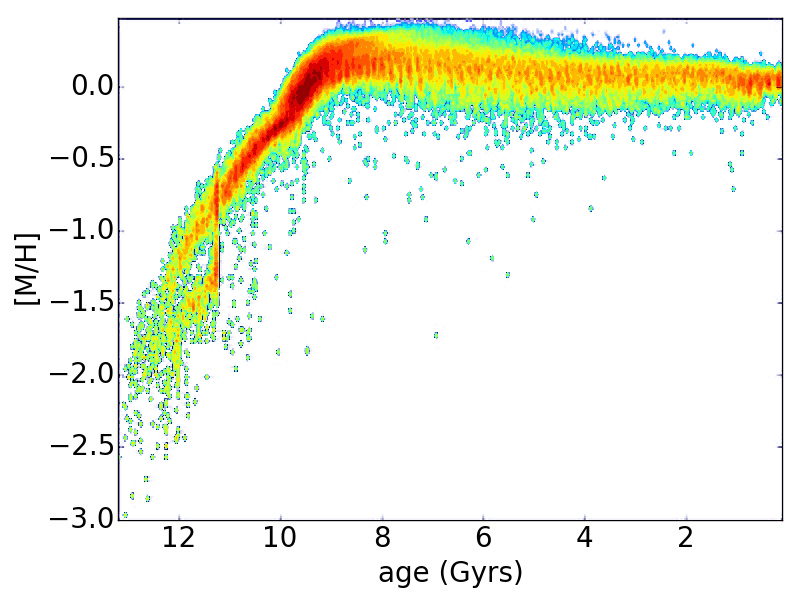}
\caption{The age-metallicity relation for stars in our 'solar region'. An early significant merger is evident around 11 Gyrs ago, followed by relatively few minor accretion events. A flat age-metallicity relation is found over the past 8 Gyrs, which corresponds to the epoch of thin disc formation and evolution. }
\label{fig:age} 
\end{figure}

Before proceeding, a quick note on terminology. Some studies refer to 
"star formation timescales", a term we find ambiguous. Does it mean 
duration of star formation, or the timescale for exhausting the existing 
gas? Here we refer to duration of star formation and star formation 
efficiency.  The paper is organised as follows: in \S2 we 
introduce the simulated Milky Way analogue; in \S3 we analyse the 
simulation and show that the different abundances in the accreted galaxy 
and the main progenitor is caused by their different star formation 
efficiencies and star formation histories; 
in \S4 we summarise and discuss the results.

\section{Simulation} \label{sec:simulations}

\subsection{Simulation details}\label{sec:simu}

The simulated Milky Way analogue galaxy used in this study comes from 
the MaGICC project \citep{brook12b,stinson12} which were the first 
cosmological hydrodynamical simulations to reproduce galaxy scaling 
relations. The simulation employs the SPH code Gasoline 
\citep{wadsley04, keller14} that includes ultraviolet (UV) heating, 
ionization and cooling due to hydrogen, helium, and metals \citep{shen10} 
and a subgrid model for turbulent mixing of metals and energy 
\citep{wadsley08}. 

Stars form from cool ($T$ $<$ $15,000$ K), dense ($n_{th}$ $>$ $10.3$~cm$^{-3}$) gas.  As metal cooling readily produces dense gas, the star formation density threshold is set to the maximum density at which gravitational instabilities can be resolved, $\frac{32 M_{gas}}{\epsilon^3}$($n_{th} > 9.3$ cm$^{-3}$), where $M_{gas}=2.2\times10^5$ M$_\odot$ and $\epsilon$ is the gravitational softening (310 pc).  Such gas is converted to stars according to the equation
\begin{equation}
\frac{\Delta M_\star}{\Delta t} = c_\star \frac{M_{gas}}{t_{dyn}}
\end{equation}
where, $\Delta M_\star$ is the mass of the star particle formed, $\Delta t$ is the timestep between star formation events ($8\times10^5$ yr in these simulations), and $t_{dyn}$ is the gas particle's dynamical time.  $c_\star$ is the efficiency parameter, i.e. the fraction of gas that will be converted into stars during $t_{dyn}$. Note that this  efficiency parameter is not to be confused with the effective star formation efficiency we refer to in this study, which is measured on larger temporal and spatial  scales, and which is largely regulated by feedback rather than by the details of the star formation implementation.

Star particles represent co-eval represent stellar populations, and the lifetimes of the constituent stars can be traced, such that they feed energy back into the ISM via 
blast-wave supernova (SN) feedback \citep{stinson06} and early stellar 
feedback from massive stars \citep{stinson13}. The AHF halo finder 
\citep{Knollmann09} is used to identify halos and the \textit{pynbody} 
package \citep{Pontzen13} is used for parts of the analysis.

Details of the chemical evolution model are found in \cite{stinson06}. 
We employ a Chabrier \citep{chabrier03} initial mass function (IMF). We use 
the \cite{raiteri96} parameterisation of stellar lifetimes for stars of 
varying metallicities. Stars with masses in the range of 8-40~M$_{\rm 
\sun}$ explode as SNeII. The number of SNeIa follows the 
\cite{raiteri96} implementation of the \cite{greggio83} 
single-degenerate progenitor model.  This ensures a finite time delay 
for the main channels of Fe production, i.e., SNeII and SNeIa.  We 
employed yields from the literature for SNeII \citep{woosley95} and 
SNeIa \citep{nomoto97}. In what follows, we 
use oxygen as a proxy for $\alpha$. These are 
relatively simplistic models, but they do allow us to identify the main 
drivers of chemical abundance evolution within the simulations.

The Numerical Investigation of a Hundred Astrophysical Objects (NIHAO, 
\citealt{wang15,obreja16}), simulated 125 galaxies using a very similar 
framework to MaGICC with some technical updates that do not result in 
significant differences in the resultant galaxies. This has allowed 
comparison of these simulations with an even larger range of 
observations \cite[e.g.][]{obreja19}, providing greater confidence in 
the adopted model. 

When selecting the Milky Way analogue for this study, 18 simulations of similar mass as the Milky Way were explored, 5 from the MaGICC project and 13 from NIHAO. We used the closest analogue to the Milky Way  (galaxy \tt g15784 \rm from the MaGICC program - see also, \citealt{gibson13,walker14a})
 in terms of the merger history, the metallicity-age relation in the solar region (see Fig 1.) and tracks of the alpha to iron abundances  in the solar region (see Fig 2.).  An  exploration of how merger history affects abundances in the discs of Milky Way analogues in the NIHAO sample can be found in \cite{buck20}.
The thin and thick discs of this particular simulated Milky Way analogue were extensively studied in  \citep{gibson13} and \citep{miranda16}, respectively, and shown to have a range of properties, in particular chemical abundance gradients, that compare favourably with observations of the Milky Way, and indeed was shown to better reproduce these than other Milky Way mass galaxies explored in those studies. It is likely no co-incidence that the Milky Way analogue that best matches the  Milky Way chemical properties has a merger history that appears to be similar to that of the Milky Way, i.e. an early significant/major merger around 10-11 Gyrs ago \citep{helmi18,chaplin20} followed by relatively  insignificant mergers and interactions which have only disturbed the disc in a minor (though interesting) way, in particular Sagittarius (\citealt{purcell11}; \citealt{laporte19}; \citealt{antoja18}; Ruiz-Lara et al. 2020).

The simulation was not tailored to have a merger that mimics that of 
GES; we simply chose a case that was the best analogue to such a 
system.
Therefore, comparisons with observations can not be expected to 
match precisely, particularly when given that the merger occurred at a 
slightly different time (most likely slightly earlier in the simulation, 
although a degree of uncertainty remains as to the exact time of the real 
merger), and with slightly different mass ratios, and that there are 
other uncertainties such as the input yields that we use in our model 
that sets the quantitive values of the ratio [$\alpha$/Fe] for example.
 
\begin{figure}
\includegraphics[width=3.4in,height=2.5in]{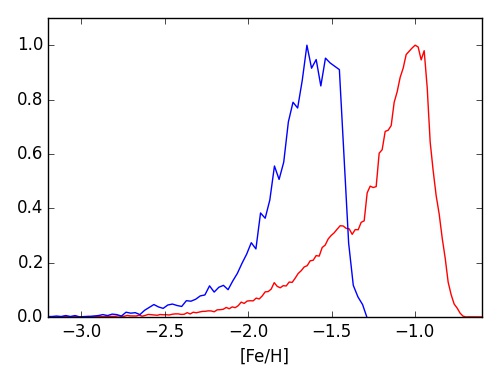}
\caption{The metallicity distribution functions of the main progenitor (red) and accreted GES analoge (blue), prior to their merger at z$\sim$2.6. }
\label{fig:MDF} 
\end{figure}

Nevertheless, the merger  history  and similarity of its disc properties to the 
observed Milky Way makes this a particularly good analogue for our 
study.  

The simulated Milky Way is the one that qualitatively best resembles the 
real Milky Way in the [$\alpha$/Fe]-[Fe/H] plane. Figure 1 shows the age-metallicity relation for stars in the 'solar region'. An early significant merger is evident around 11 Gyrs ago, redshift $z$$\sim$2.6,  followed by relatively few minor accretion events. A flat age-metallicity relation is found over the past 8 Gyrs, which corresponds to the epoch of thin disc formation and evolution.

\subsection{Properties of the main progenitor and satellite prior to the merger}
 
 We measure the masses 
of the main progenitor and accreted galaxy prior to the latter
entering the virial radius of the former, meaning that the 
total, stellar, and gas masses of the merging galaxies shown in 
Table~\ref{tab:mass} are measured at redshift $z$$\sim$3.
Figure~\ref{fig:MDF} shows the significantly different metallicity distribution functions of the main progenitor (red) and GES analogue (blue) prior to their merger. Both the main progenitor and the GES analogues are around 0.2-0.3 dex lower in metallicity than the corresponding populations observed in the Milky Way \cite[see e.g. Fig 2 of][]{gallart19}. From the slope of the age-metallicity relation in Fig. 1, one can see that the metallicities would match the observed ones if the merger was around 0.5 Gyrs later, i.e.  around 10.5 Grys ago rather than 11 Gyrs ago as happened in this simulation. A systematic offsets between the metallicities in the simulation and observations may also be attributed, in part, to our specific choice of yields and IMF. 

\begin{table}
\label{tab:masses}
\begin{center}
\begin{tabular}{lccc}
\hline
\hline
 & M$_{\rm total}$ & M$_{\rm stars}$   & M$_{\rm gas}$\\
\hline
Main Progenitor   & 2.3e11 & 2.0e9 & 3.4e10   \\                    
Accreted Galaxy   & 9.7e10 & 2.9e8 & 1.5e10  \\                    
\hline
\hline
\end{tabular}
\caption{Total, stellar and gas mass of the main progenitor and accreted gaia-Enceladus sausage analogue, prior to their merger.}
\label{tab:mass}
\end{center}
\end{table}

\begin{figure}  
\hspace{.0cm}\includegraphics[width=3.4in,height=2.5in]{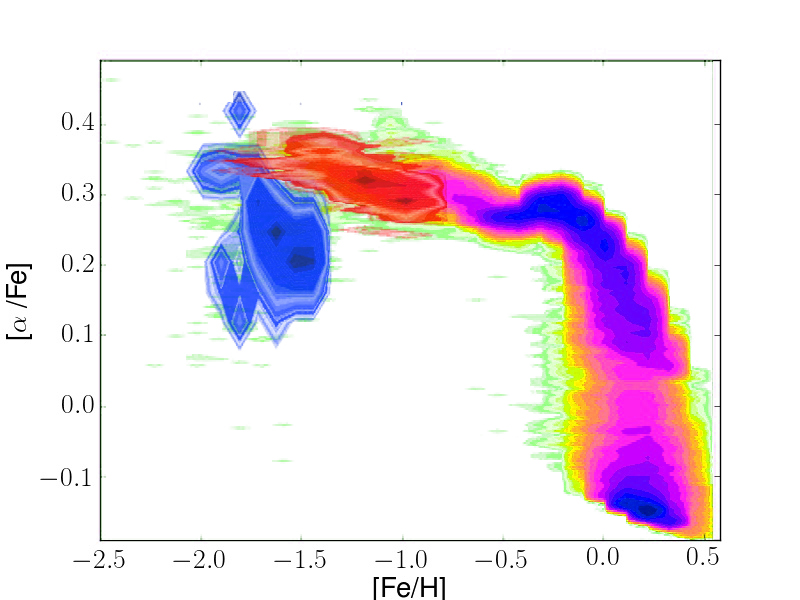}
\caption{[$\alpha$/Fe] vs [Fe/H] of the 'solar neighbourhood' region in 
the simulated galaxy, defined as an annulus of 7-9 kpc from the 
galactic centre and within 2kpc of the disc plane. The thick and 
thin disc separation is evident at [$\alpha$/Fe]$\sim$0 [Fe/H]$\sim$0. 
Overlaid are the stars that originated in the main progenitor (red) and 
major accreted satellite (blue) prior to their merger at z$\sim$2. Their 
different tracks in this plane are evident; it is also evident that the 
main progenitor smoothly blends into the ongoing thick disc track.}
\label{fig:alphafe} 
\end{figure}

\subsection{The [$\alpha$/Fe] vs [Fe/H] plane }\label{sec:global}

Figure~\ref{fig:alphafe} shows the [$\alpha$/Fe] vs [Fe/H] for the 
'solar neighbourhood' region in the simulated galaxy.  We define the 
solar neighbourhood as within an annulus of 7-9 kpc from the galactic 
centre and within 2kpc of the plane of the disc. This is not quite the 
same as a 2kpc sphere volume around the Sun in the Galaxy, but allows us 
to sample a similar region that has a larger volume, which is required 
as the simulation has far fewer 'star particles' compared to the 
observed {\it Gaia} sample of stars. The results are robust to the precise region that we chose as the Solar Neighbourhood, with our main conclusions not affected by chosing any region from 4-11 kpc in the disc within 4 kpc of the disc plane.

The thick and thin disc separation 
is evident at [$\alpha$/Fe]$\sim$0 [Fe/H]$\sim$0. Overlaid are the stars 
that originated in the main progenitor (red) and major accreted 
satellite - i.e., the GES analogue (blue) prior to their merger at 
z$\sim$2. The different tracks in this plane are evident. It is also 
evident that the main progenitor smoothly blends into the thick disc 
track, which continues to form after the merger. The aim of this paper 
is to better understand the physics driving these different tracks,
and their trajectories, for 
the main progenitor and the accreted GES analogue galaxy, prior to their 
merger.

\section{Results: Explaining the  trajectories in the [$\alpha$/F{\tiny e}] vs [F{\tiny e}/H]  plane} \label{sec:results}

\subsection{Global properties of main progenitor and accreted galaxy}

Table~\ref{tab:mass} shows the total, stellar and gas masses of the two 
merging galaxies. The main progenitor is 2.4 times more massive than the 
accreted galaxy, but has a factor of 6.9 more mass in stars. This is in 
agreement with the stellar mass-halo mass relation for galaxies at 
high-redshift \citep{behroozi13}. Yet the ratio of the gas masses (2.3) 
follows closely the ratio of total masses rather than the ratio of stellar 
masses. This also makes sense, with lower mass galaxies known to have 
higher gas fractions than higher mass galaxies in this mass range, at 
least at $z$=0 \citep[e.g.][]{bradford15}.

In what follows, we define star formation efficiency (SFE) as the rate at which gas is converted to stars, i.e. 
SFE$=$SFR/M$_{gas}$
where SFR is the star formation rate and M$_{gas}$ is the total gas mass within the virial radius. Note that SFE is defined as the ratio of star formation rate 
and total gas mass within the virial radius, rather than cold gas or HI 
gas. This allows the accounting for ongoing gas cooling, in 
particular hydrogen rich gas, into the star forming regions, as well as 
the outflow of gas beyond the star forming regions. We also  explore the effect of restricting to observable gas.

\begin{figure}
\includegraphics[width=3.4in,height=5in]{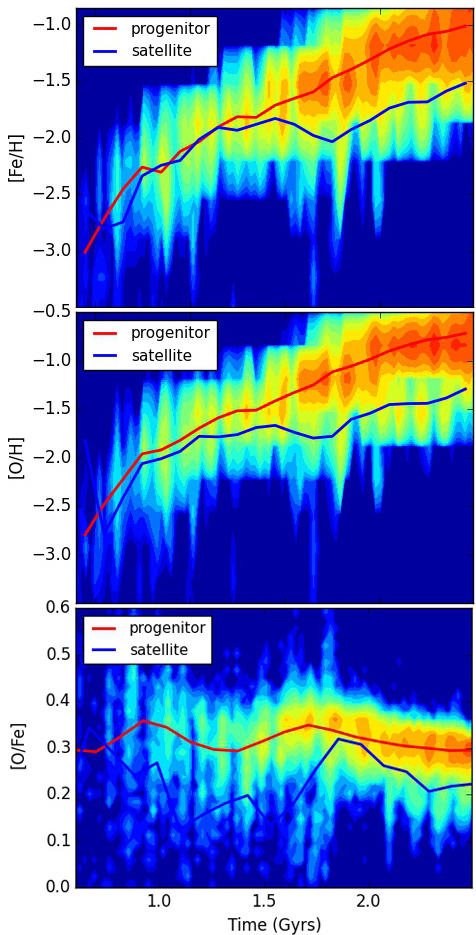}
\caption{{\it Top panel}: The background density shows the time 
evolution of [Fe/H] for all 'solar neighbourhood' stars in the final 
galaxy. Evolution of the mean values for the stars that formed before 
the merger and originated in the main progenitor (red) and the accreted 
galaxy (blue) are shown as lines. The two galaxies are seen to have different age-metallicity relations.
{\it Middle panel}: Same as the top panel, but for [O/H].
{\it Bottom panel}:  Same as the top panel, but for [O/Fe].
}
\label{fig:abun} 
\end{figure}

Overall, the merging galaxies have very similar baryon fractions within 
their virial radii. This implies that preferential outflows of metals in 
the lower mass galaxy is not the driver of the different tracks in 
[$\alpha$/Fe] space.  However, the two merging galaxies do have 
different efficiencies of converting gas into stars. Could a difference 
in star formation efficiency be driving the different tracks in the 
[$\alpha$/Fe] vs [Fe/H] plane?

\begin{figure}  
\hspace{.0cm}\includegraphics[width=3.4in,height=2.5in]{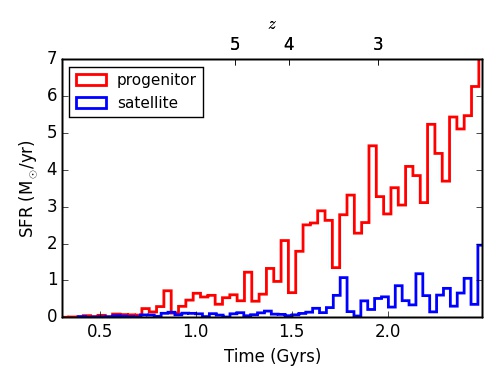}
\caption{The star formation histories of the main progenitor (red) and 
'to be accreted' GES-like galaxy (blue) prior to the accreting galaxy 
entering the virial radius of the main progenitor. The star formation 
rate is increasing for the main progenitor, whilst it is almost constant 
and regulated by feedback in the case of the lower mass accreting 
galaxy.}
\label{fig:sfh} 
\end{figure}

\subsection{Evolution of [Fe/H], [O/H] and [Fe/O]}\label{sec:sfh}

We look for clues in the temporal evolution of [Fe/H], [O/H], and [O/Fe].
The top panel of Figure~\ref{fig:abun} shows 
that the main progenitor reaches a factor of $\sim$3 higher in 
[Fe/H] than does 
the accreted galaxy by the time of the merger. Similarly, the middle 
panel of Figure~\ref{fig:abun} shows the same behaviour, both
qualitatively and quantitatively, for [O/H]. In other words, the ISM
is more enriched in both iron and oxygen in the more massive progenitor, 
a natural consequence of having converted a higher fraction of its
baryons into stars.
 
By contrast, the difference in [O/Fe] is relatively small (no more than
$\sim$0.1~dex, as per the 
bottom panel of Figure~\ref{fig:abun}. One can then 
see that the highest density red region in Figure~\ref{fig:alphafe} is 
offset approximately 0.5 dex rightwards and 0.1 dex upwards from the 
highest density blue region.  We attribute the 0.1 dex offset to the 
difference in the rate of change of star formation. In particular, the 
main progenitor has an {\it increasing} star formation rate as we show 
next.

\subsection{Star Formation Histories }\label{sec:sfh}
 
In Figure~\ref{fig:sfh}, we show the star formation histories of the main 
progenitor (red) and accreted galaxy (blue) prior to the accreted galaxy 
entering the virial radius of the main progenitor. It is clear that the 
star formation rate is rapidly increasing for the main progenitor. By 
contrast, feedback is able to regulate the star formation and keep the 
relatively constant star formation rate in the lower accreting mass 
galaxy. An increasing star formation rate can sustain a relatively high 
[$\alpha$/Fe] in the main progenitor, because the rate of SNeII is 
linked to the current SFR, whilst the rate of SNeIa is related to the 
earlier (and lower) SFR. Once the star formation rate peaks and starts 
dropping, this situation is reversed: the rate of SNeIa relates to the 
earlier (and higher) SFR whilst the rate of SNeII is determined by the 
current (lower) SFR. This transition from increasing to decreasing star 
formation rates could be expected to result in a relatively rapid 
transition from high to low [$\alpha$/Fe].
 
Looking at the [Mg/Fe]-[Fe/H] plane of data from APOGEE DR14, as shown 
in Figure~1 of \cite{mackereth18}, one can see that similar offsets of 
0.5 dex rightwards and 0.1 dex upwards, as we found in our simulations, 
can explain the observational results.

\begin{figure}
\includegraphics[width=3.4in,height=2.5in]{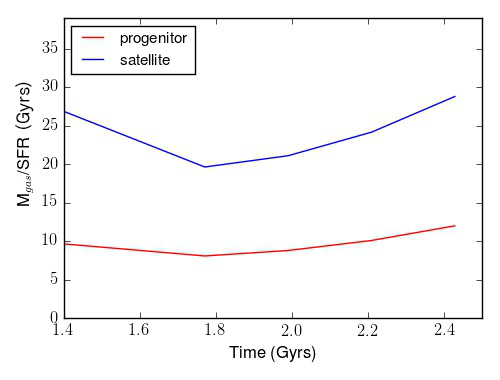}
\caption{The inverse of the star formation efficiencies, i.e. total gas 
mass/star formation rate, for the main progenitor and the accreted 
satellite, leading up to their {\large }merger in the simulation. }
\label{fig:sfe} 
\end{figure}

\subsection{Star Formation Efficiencies}

Because the Kennicutt-Schmidt law has a slope greater than 1, an 
increased star formation rate is associated with an increased star 
formation efficiency. In the simulations, where a Kennicutt-Schmidt law 
is imposed, this is also reflected in the greater star-to-gas ratio in 
the main progenitor than in the accreted galaxy, as shown in 
Table~\ref{tab:mass}. In Figure~\ref{fig:sfe} we show the inverse of the 
star formation efficiencies (SFEs) of the main progenitor and the 
accreted satellite, during the time leading up to their merger in the 
simulation.  The plot shows that 
the main progenitor would consume its gas within ten Gyrs, whilst the 
satellite would take 25 Gyrs, more than a Hubble time, to consume its 
gas.

When restricting to only include `observable gas' or gas in the star 
forming region, meaning 4/3 times the HI gas mass, the SFE is around an 
order of magnitude higher for both the main progenitor and the accreting 
galaxy, increasing from $\sim$0.04 and $\sim$0.1 to $\sim$0.4 and 
$\sim$1 Gyrs$^{-1}$ respectively. The difference in star formation 
efficiency remains evident when the definition of star formation 
efficiency is restricted to 'observable' gas.

\subsection{Gas density profiles}

In Figure~\ref{fig:den} we plot the density profile of gas for 
the main progenitor (red) and accreted satellite (blue) prior to their 
merger. The higher density in the inner star forming region of the main 
progenitor drives higher star formation rates which, via the (imposed) 
Kennicutt-Schmidt law, means higher star formation efficiencies. We note 
as well the slight differences in the shape of the profiles.

\vspace{-.5cm}\section{Conclusions} \label{sec:discussion}

Using a simulated Milky Way analogue galaxy that has a final significant 
merger (major merger in terms of total mass ratios) at early times, qualitatively mimicing the Gaia-Encaladas-Sausage (GES) accretion event, we have 
explored the cause of the different abundances of the high velocity 
(halo) stars that were accreted compared to those that were heated from 
the main progenitor (this 'in situ' halo population has also been 
referred to as the high velocity tail of the thick disc). 
We show that, prior to this early merger, the main progenitor and merging GES analogue have different star formation histories and different age-metallicity relations.  

Our main 
finding is that the difference in abundances is driven by  different 
star formation efficiencies. Possible other explanations that are not supported by our 
model include different durations of star formation, preferential 
outflows of particular metals, or different Initial Mass Functions 
(IMF). We discuss them in turn below:

\begin{figure}
\includegraphics[width=3.4in,height=2.5in]{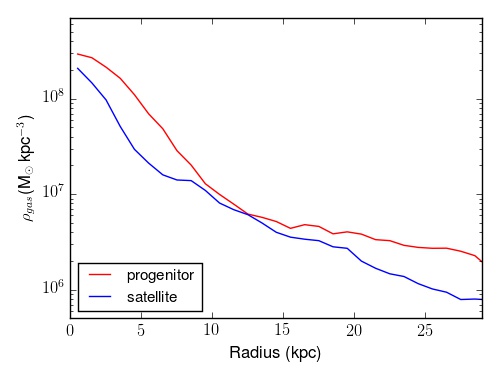}
\caption{The density profile of gas for the main progenitor (red) and 
accreted satellite (blue) prior to their merger.}
\label{fig:den} 
\end{figure}

The finding of common ages (\citealt{gallart19}, see also 
\citealt[][]{sahlholdt19}) for halo stars formed in GES and the main 
progenitor allowed us to rule out a longer duration of star formation as 
an explanation for the two different tracks these populations take in 
the [Fe/H] vs [$\alpha$/Fe] plane. This degeneracy breaking allowed us 
to explore the origin of these differing tracks in abundance space.

We found that the main progenitor and accreted galaxy retain the same 
amount of baryons prior to their merging, allowing us to rule out 
outflows as the main driver of the different tracks they take through 
the [Fe/H] vs [$\alpha$/Fe] plane.

Finally, our study does not rule out the possibility of a varying IMF in 
the different galaxies that formed the bulk of the inner halo. Some 
degenerate model may be possible in which the IMF in the more massive 
galaxies have more massive stars than the less massive galaxies. Such a 
model would need to match the vast array of galaxy properties that our 
simulations have been shown to reproduce. We simply have shown that a 
varying IMF is not required, at least in these relatively low mass 
systems that merged around 10 Gyrs ago.

We note that, importantly, the simulated Milky Way shares many features 
of the observed Milky Way, particularly in its halo, thick disc and thin 
disc kinematics and abundances \citep[see][as well as Fig.~1 of this 
work]{brook12b,gibson13,walker14,miranda16,gallart19}. We also 
note that this same model 
for galaxy formation reproduces observed galaxy scaling relations over a 
wide range of masses. This is important because this again breaks many 
degeneracies that may exist in models that purely match the relation of 
interest in this paper - i.e., models that may reproduce the [Fe/H] vs 
[O/Fe] features but cannot form realistic galaxies in a broader context.

Our explanation is the most simple one in many regards. The 
Kennicutt-Schmidt law that relates gas density to star formation 
has a power law index greater than unity, which implies that galaxies with 
higher star formation rates have higher star formation efficiencies. The 
lower star formation efficiencies in the lower mass galaxy means that 
gas remains generally metal-poor, even with the two galaxies forming 
stars over the same length of time. In addition, the {\it increasing} 
SFR in the main progenitor helps in keeping the relatively high [O/Fe], 
by ensuring that SNII yields remain dominant over SN1a. We believe that 
this is the dominant mechanism driving the different tracks in the 
[Fe/H] vs [O/Fe] between GES and the main progenitor. Combined, these 
mechanisms lead to the different chemical evolution paths between the 
main progenitor and GES before the merger, even with the two galaxies 
forming stars over the same length of time.
 
This is an important finding as it is only in the Milky Way that such 
detailed abundances can be studied and related to early forming 
galaxies. It is quite possible that our explanation can be generalised 
to the origin of the different [Fe/H] vs [O/Fe] relations for different 
galaxies and/or components.

Although other explanations are possible in the parameter space 
provided, we argue that the one presented here gains support from the 
self-consistent cosmological simulation model from which it was drawn 
naturally. That is to say, cosmological simulations using the same 
physical recipes have been shown to match a wide range of observed 
galaxy properties over a wide range of galaxy masses. This is 
particularly important when exploring accretion of low mass objects onto 
Milky Way mass galaxies: it is crucial to be able to reproduce the 
properties of lower mass galaxies in order to have confidence in the 
properties of the accreted galaxy.

\vspace{-.6cm}\section*{Acknowledgements}
BKG acknowledges the support of STFC through the University of Hull 
Consolidated Grant ST/R000840/1 and access to {\sc viper}, the University 
of Hull High Performance Computing Facility.

\vspace{.0cm}\bibliographystyle{mn2e}
\bibliography{archive}


\label{lastpage}

\end{document}